\documentclass[sigconf]{acmart}

\def\BibTeX{{\rm B\kern-.05em{\sc i\kern-.025em b}\kern-.08emT\kern-.1667em\lower.7ex\hbox{E}\kern-.125emX}}

\usepackage{soul} % AGGIUNTO DA LILIANA per fare stright through
\usepackage{graphicx}
\usepackage{balance}

\copyrightyear{2019} 
\acmYear{2019} 
\setcopyright{acmlicensed}
\acmConference[UMAP '19]{27th Conference on User Modeling, Adaptation and Personalization}{June 9--12, 2019}{Larnaca, Cyprus}
\acmPrice{15.00}
\acmDOI{10.1145/3320435.3320441}
\acmISBN{978-1-4503-6021-0/19/06}

\begin{document}
\title{
Multi-faceted Trust-based Collaborative Filtering
}

\author{Noemi Mauro}
 \orcid{}
 \affiliation{%
   \institution{Computer Science Department University of Torino}
   \streetaddress{Corso Svizzera 185}
   \city{Torino} 
   \state{Italy} 
   \postcode{10149}
 }
\email{noemi.mauro@unito.it}

\author{Liliana Ardissono}
 \orcid{}
 \affiliation{%
   \institution{Computer Science Department University of Torino}
   \streetaddress{Corso Svizzera 185}
   \city{Torino} 
   \state{Italy} 
   \postcode{10149}
 }
\email{liliana.ardissono@unito.it}

\author{Zhongli Filippo Hu}
 \orcid{}
 \affiliation{%
   \institution{Computer Science Department University of Torino}
   \streetaddress{Corso Svizzera 185}
   \city{Torino} 
   \state{Italy} 
   \postcode{10149}
 }
\email{zhongli.hu@edu.unito.it}

\renewcommand{\shortauthors}{}

\begin{abstract}
Many collaborative recommender systems leverage social correlation theories to improve suggestion performance. However, they focus on explicit relations between users and they leave out other types of information that can contribute to determine users' global reputation; e.g., public recognition of reviewers' quality.

We are interested in understanding if and when these additional types of feedback improve Top-N recommendation.
For this purpose, we propose a multi-faceted trust model to integrate local trust, represented by social links, with various types of global trust evidence provided by social networks. We aim at identifying general classes of data in order to make our model applicable to different case studies.
Then, we test the model by applying it to a variant of User-to-User Collaborative filtering (U2UCF) which supports the fusion of rating similarity, local trust derived from social relations, and multi-faceted reputation for rating prediction. 

We test our model on two datasets: the Yelp one publishes generic friend relations between users but provides different types of trust feedback, including user profile endorsements. The LibraryThing dataset offers fewer types of feedback but it provides more selective friend relations aimed at content sharing. The results of our experiments show that, on the Yelp dataset, our model outperforms both U2UCF and state-of-the-art trust-based recommenders that only use rating similarity and social relations. Differently, in the LibraryThing dataset, the combination of social relations and rating similarity achieves the best results.
The lesson we learn is that multi-faceted trust can be a valuable type of information for recommendation. However, before using it in an application domain, an analysis of the type and amount of available trust evidence has to be done to assess its real impact on recommendation performance.

% LINK AI CCS CONCEPTS: \url{https://dl.acm.org/ccs/ccs_flat.cfm#10002951}

\end{abstract}
% \category{H.3.3}{Information Search and Retrieval}{Information filtering} \category{See
%   \url{http://acm.org/about/class/1998/} for the full list of ACM
%   classifiers. This section is required.}{https://www.acm.org/publications/computing-classification-system/1998/ccs98}{}


%\printccsdesc
%\keywords{\plainkeywords}

\maketitle

\section{Introduction}

Social correlation theories, such as homophily \cite{McPherson-etal:01} and social influence \cite{Marsden-Friedkin:93}, associate social links (e.g., friends relations in human networks) to user similarity by observing that ``similarity breeds connection" \cite{McPherson-etal:01}; moreover, homophily has been recently observed in digital social networks \cite{Aiello-etal:12}. Building on these findings, researchers developed several collaborative recommender systems that employ friends, follower and trust links between users, in combination with rating similarity, to improve the estimation of user preferences in cold start scenarios. These systems assume that social relations are associated with user trust, or with pseudo-trust, as in friend networks. Moreover, they predict ratings by relying on local trust that is attributed to the people directly linked to the current user, or reachable through a short path of links; e.g., \cite{Ma-etal:11,Jamali-Ester:10,Yang-etal:17}.
In \cite{Tang-etal:13}, Tang et al. observe that, in the physical world, people are likely to ask their local friends for suggestions, but they also tend to seek advice from reputable users. Therefore, in the LOCABAL recommender system, the authors combine local trust with global user reputation; however, they compute reputation on the sole basis of social links by applying the PageRank \cite{Page-etal:99} metric.
We point out that additional types of trust statements could be considered to build a richer reputation model. For instance, various online location-based services such as Yelp \cite{Yelp}, Booking \cite{Booking} and Expedia \cite{Expedia}, enable users to provide feedback on other users and on item reviews. These systems do not disclose the identity of the people who provide the feedback; however, they publish aggregated data that can be interpreted as ``anonymous trust statements''. 
We are interested in evaluating the impact of this type of information on Top-N recommendation.

For this purpose, we propose a multi-faceted trust model that integrates local trust with global reputation evidence available in social networks and e-commerce sites. We define four general classes of evidence, which can be mapped to different types of information published by social networks. Specifically, we estimate the multi-faceted global reputation of a user by analyzing the trust statements provided by the other users; e.g., endorsements to her/his public profile and feedback on item reviews. 
Moreover, we model local trust between users by taking into account both direct friends and the relations depending on the existence of implicit user groups, which can be revealed by the presence of relevant numbers of common friends and might denote preference similarity.

We test our model on collaborative recommendation by applying it to a variant of User-to-User Collaborative Filtering (U2UCF) which we denote as Multi-faceted Trust-based Recommender (MTR). MTR can be configured to combine in different ways rating similarity, local trust derived from social relations and multi-faceted global reputation for neighbor detection and rating estimation. MTR also makes it possible to separately include or exclude different facets of trust; thus, it helps understand their impact on accuracy.
We compare MTR with U2UCF, and with LOCABAL and TrustMF trust-based recommender systems, which combine rating similarity and trust information in Matrix Factorization.

For the experiments, we work on two datasets: the Yelp one \cite{Yelp-dataset} provides information about generic friend relations established by users to filter contributions in the social network. The dataset publishes different types of trust feedback, including user profile endorsements. The LibraryThing dataset \cite{LibraryThing-Dataset} stores fewer types of feedback but it publishes more selective friend relations aimed at content sharing. The evaluation provides the following results:
\begin{itemize}
    \item 
    On the Yelp dataset, multi-faceted trust information helps recommendation performance, probably by complementing the social ties defined by friend relations. The MTR configurations that combine social relations and global multi-faceted reputation outperform U2UCF in accuracy, MRR and diversity of recommendations. Moreover, they outperform LOCABAL and TrustMF in RMSE and MAE. Furthermore, MTR obtains the best accuracy when local trust includes both direct social links and implicit groups of users having a relevant number of common friends. 
    In contrast, in the LibraryThing dataset, LOCABAL obtains the best performance results by inferring users' reputation on the basis of friend relations, and by combining it with rating similarity.
    \item 
    Profile endorsements and feedback on users' contributions improve recommendation performance: the configurations of MTR ignoring these types of information (and especially the former) have lower accuracy, MRR and diversity than the other MTR configurations. However, taking profile endorsements into account reduces user coverage.
\end{itemize}
We conclude that multi-faceted trust can help recommendation, especially when social relations are weak trust predictors, because it complements them with global reputation data. However, before using this type of information in a social network, an analysis of the type and amount of available trust evidence, as well as of the meaning of social relations, is needed to assess its real impact on recommendation performance. 
In summary, this paper provides the following contributions:
\begin{itemize}
    \item A novel, multi-faceted model of trust between users in a social network that includes local trust relations and global reputation gained on the basis of different types of feedback; e.g., user profile endorsements and appreciations of user contributions such as item reviews.
    \item A new definition of local trust aimed at considering both direct friends and users having a relevant number of common friends, as this type of information helps identify implicit user groups.
    \item Experimental results aimed at understanding the impact of different trust facets on recommendation performance in Collaborative Filtering. 
\end{itemize}
The remainder of this paper is organized as follows: Section \ref{related} positions our work in the related literature. Section \ref{model} presents our multi-faceted trust model and its integration in Collaborative Filtering. Section \ref{test} describes the evaluation methodology and results. Finally, Section \ref{conclusions} concludes the paper and outlines our future work.

\section{Background and related work}
\label{related}

\subsection{Basic Concepts: Trust and Reputation}
%The notion of {\em trust} has been described in different ways depending on the application scenarios. 
According to Gambetta \cite{Gambetta:88}, ``trust [...] is a particular level of the subjective probability with which an agent will perform a particular action, both before [we] can monitor such action (or independently of his capacity of ever to be able to monitor it) and in a context in which it affects [our] own action.'' Moreover, Misztal points out that trust is a subjective degree of belief about agents'' \cite{Misztal:96}.
Furthermore, Goldbeck and Hendler specify that a user trusts another one in a social network if (s)he believes that any future transaction with her/him will be rewarding rather than detrimental \cite{Goldbeck-Hendler:04}.

Differently, {\em reputation} ``helps us to manage the complexity of social life by singling out trustworthy people - in whose interest it is to meet promises'' \cite{Misztal:96}. In \cite{Abdul-Rahman-Hailes:00} reputation is described as ``an expectation about an agent's behaviour based on information about or observations of its past behaviour''.
In \cite{McNally-etal:14}, McNally et al. point out that reputation can derive from direct user to user interaction (e.g., when users are rated) or from indirect one (e.g., when they interact by virtue of some item). Moreover, it can derive from explicit trust statements (e.g., ratings) or from implicit ones such as follower relations. Overall, reputation represents a global point of view about a user.

In \cite{Josang-etal:07,Josang-etal:09}, J{\o}sang et al. discuss that Trust and Reputation Systems are challenged by strategic manipulation and by various types of attacks which cannot always be detected by statistical analyses. Therefore, in \cite{Josang:12} the authors highlight the importance of strengthening legislation as a barrier to discourage malicious behavior.

\subsection{Trust-based Recommender Systems} 
Most trust-based recommenders leverage social influence to estimate ratings in Collaborative Filtering; see \cite{Richthammer-etal:17}. They assume that the trust relations between specific users can be inferred from social links; e.g., from friend associations and/or follower relations  \cite{Goldbeck-Hendler:04,Massa-Avesani:07,Groh-Ehmig:07,Liu-Lee:10,Jamali-Ester:10,Liu-Aberer:13,Yang-etal:12,Tang-etal:13,Yang-etal:17,Ardissono-etal:17c}. McNally et al. generalize trust relations by analyzing the occurrence of collaboration events involving users \cite{McNally-etal:14}. All these works focus on the known social links existing among individual users. In comparison, we propose a multi-faceted trust model that also takes into account anonymous feedback about contributions, and endorsements of user profiles, commonly available in social networks. Furthermore, we analyze the impact of different classes of trust evidence on recommendation performance.
Finally, while in \cite{Victor-etal:11,Rafailidis-Crestani:17} both trust and distrust are considered, we focus on the former because it is more available than distrust information in social network datasets.

In \cite{Guo-etal:15}, \cite{Ma-etal:11} and \cite{Li-Fang:18}, the authors discuss that, different from explicit trust relations (such as those among Epinions users \cite{ePinions-dataset}), friendship does not strictly imply preference similarity: user preferences are strongly correlated among trusted neighbors, but only slightly positively correlated among ``trust-alike'' neighbors, such as friends in social networks \cite{Guo-etal:15}.
Moreover, several authors recognize the importance of limiting the social context to the local proximity of the user. For instance, \cite{Massa-Avesani:07} and \cite{Yuan-etal:11} prove that, when indirect connections are used to estimate preferences, recommendation accuracy sensibly decreases. In order to address this issue, various types of social regularization combine the rating and trust matrices to focus on users having preferences similar to those of the current user; e.g., via matrix factorization in \cite{Ma-etal:11,Jiang-etal:12,Ma-etal:11b,Liu-Aberer:13} or co-clustering in \cite{Du-etal:17}. 
In \cite{Guo-etal:15}, Guo et al. propose TrustSVD that extends SVD++ \cite{Koren:08} to jointly factorize the ratings and trust matrices: they learn a truster model describing how people are influenced by their parties. TrustMF \cite{Yang-etal:17} learns both the truster and trustee models because, in a social context, people mutually influence each other. 
In comparison, our model fuses social relations with multi-faceted global reputation derived from different types of feedback about users and their contributions. 

Our work takes inspiration from LOCABAL \cite{Tang-etal:13} in fusing local trust and global reputation.
However, we combine these two types of information in a K-Nearest Neighbor approach and we employ the local recommendation context given by friend relations, together with a multi-faceted global reputation derived from various types of anonymous trust statements. Moreover, we introduce a social proximity metric that takes direct friends and implicit user groups into account for neighborhood identification.

Some collaborative recommender systems integrate multiple trust relations for rating estimation; e.g., references in research papers and bookmarks in social reference management systems \cite{Alotaibi-Vassileva:16}. Moreover, in \cite{Yuan-etal:11} friendship and group membership relations are fused, taking into account that the latter denote interest similarity. Furthermore, \cite{Konstas-etal:09} uses both friend relations and tagging behavior in neighbor detection. These works exploit complementary trust information with respect to ours. Moreover, our work differs from the one in \cite{Yuan-etal:11} because we do not work on explicit user groups, but we infer implicit ones from the existence of common friends.

Reputation has also been inferred from correlation in rating behavior; e.g., (i) as the percentage of ratings provided by a user that agree with those of the other raters \cite{O'Donovan-Smyth:05}, (ii) by clustering users on the basis of their rating similarity \cite{Su-etal:17} (in this case, the ``honest'' group is the largest cluster), or (iii) by iteratively calculating the correlation of the historical ratings provided by a user and the intrinsic qualities of items emerging from the ratings they receive \cite{Qian-etal:16}. 
Our current model does not cover rating correlation because a deeper investigation of the phenomenon is needed to take into account user diversity. For instance, Victor et al. point out that controversial reviews have to be considered and matched to individual preferences  \cite{Victor-etal:11}.

\section{Recommendation Model}
\label{model}

\subsection{Multi-faceted Trust Model}
Our model is based on types of trust evidence that are publicly provided by social networks such as Yelp \cite{Yelp}, Booking \cite{Booking}, Expedia \cite{Expedia}, LibraryThing \cite{LibraryThing} and Airbnb \cite{Airbnb}.
Sections \ref{sec:YELP} and \ref{sec:LT} provide examples of the application of our model.

Let $U$ be the set of users and $I$ the set of items of a social network. Given $v\in U$ and $i \in  I$, we consider the following classes of trust evidence:
\begin{enumerate}
        \item[(A)] 
        {\em Global feedback on the user's profile:}
        \begin{itemize}
         \item {\em User profile endorsements and public recognition ($endors_v$, in [0, 1]):} this class represents the degree of appreciation that {\em v}'s public user profile has received from the users of the social network (e.g., number of ``likes'') and public assessments of reputation (e.g., number of years in the status of ``Elite'' contributor granted by the social network itself, as in Yelp). The value of this trust facet is evaluated as the ratio between the number of appreciations received by {\em v}'s profile ($Appreciations_v$) and the maximum number of appreciations received by a profile $a \in U$:
        \begin{equation}
            fEndors_v = \frac{|Appreciations_v|} 
            {max_{a \in U} |Appreciations_a|}
        \label{eq:fEndors}
        \end{equation} 
        \item {\em Visibility}: this class is aimed at estimating how popular $v$ becomes thanks to her/his contributions; e.g., item reviews and comments. This is computed as the ratio between the number of appreciations received by $v$'s profile and the total number of contributions provided by her/him ($ Contributions_v$), normalized by the maximum number of appreciations acquired by the other users:
        \begin{equation}
            vis_v = \frac{|Appreciations_v|}
            {max_{a \in U} (|Appreciations_a|*|Contributions_v|)}
        \label{eq:visibility}
        \end{equation} 
        \end{itemize}
    \item[(B)] 
    {\em Global feedback on $v$'s contributions ($fContr_v$, in [0, 1]):} this class summarizes the degree of appreciation that the contributions posted by {\em v} receive, compared to that of the other users of the social network:
         \begin{equation}
            fContr_v = \frac{ 
        	    \sum\limits_{x\in Contributions_v} |Appreciations_x|}
            {max_{a\in U} (\sum\limits_{y\in Contributions_a} |Appreciations_y|)}\\
        \end{equation} 
    \item[(C)]
    {\em Global feedback on {\em v}'s review of item} $i$ {\em ($fRev_{vi}$, in [0, 1]):} this class is aimed at promoting the authors of popular reviews. It is computed as the ratio between the feedback obtained by the specific review, $rev_{vi}$, and the maximum amount of feedback obtained by the other reviews on the same item:
    \begin{equation}
       fRev_{vi} = \frac{ 
       |Appreciations_{rev_{vi}}|
        }{max_{a\in U}|Appreciations_{rev_{ai}}|}
        \label{fRev_vi}
    \end{equation} 
    \item[(D)]
    {\em Social relation between u and v ($rel_{uv}$, in [0, 1]):} this class represents the degree of local trust that $u \in U$ has in {\em v}, given the social link existing between them. We consider two alternative models of the relation between users:
    \begin{enumerate}
        \item {\em Direct connections:} in this model, only the users directly linked to $u$ are considered as trustworthy:
            \begin{equation}
             rel_{uv} =
            \begin{cases}
                1 & \quad \text{if } u ~ \text{and} ~ v ~\text{are directly linked}\\
                0 & \quad \text{otherwise}
            \end{cases}
        \label{eq:friends}
        \end{equation}
        \item {\em Direct connections + social intersection:} in this case, both the direct connections and the users having a relatively high number of direct social connections in common with $u$ are considered as trustworthy. The rationale is that, if two users have several common friends, then they might belong to a group of people having similar interests. We capture this intuition by applying the Jaccard Similarity to $u$ and $v$'s social connections:
        \begin{equation}
             rel_{uv} =
            \begin{cases}
                1 & \quad \text{if } u ~ \text{and} ~ v ~\text{are directly linked}\\
                JS(u, v) & \quad \text{otherwise}
            \end{cases}
            \label{eq:friendsJS}
        \end{equation}
        where
        \begin{equation}
             JS(u, v) = \frac{ 
       |Friends_u \cap Friends_v|}
        {|Friends_u \cup Friends_v|}
        \end{equation}
    \end{enumerate}
\end{enumerate}
The above classes of trust evidence are generic and most of them could be mapped to multiple trust indicators in a social network. For instance, Yelp supports different types of endorsements on user profiles; e.g., ``thanks'' and ``Elite'' recognition. Moreover, in other cases, the social relation between users might be mapped to friends, follower and trust relations. 
Therefore, we need a flexible way to compose information items within a unified trust model.

We fuse individual trust indicators in a linear combination, assuming that they contribute to increasing {\em v}'s trustworthiness in an additive way. 
Let's consider a set of indicators $\{f_1, \dots, f_z\}$ that are instances of the above trust evidence classes. Then, we estimate the trust of a user {\em u} in {\em v}, in the context of item $i$, as follows:
\begin{equation}
    t_{uvi} = 
    \frac{\sum_{x=1}^{z} w_x * f_x}
    {\sum_{x=1}^{z} w_x} 
    \label{eq:hybridReputation}
\end{equation}
where $t_{uvi}, w_1, \dots, w_z$ take values in the [0, 1] interval.

We assume that each trust indicator is computed according to the method defined for the class to which it belongs. Notice that the computation of $t_{uvi}$ can be performed  by maintaining:
\begin{itemize}
\item
For each unidimensional trust indicator (e.g. $fEndors_v$, $vis_v$, $fContr_v$), a vector of length |U|;
\item
A $U X I$ matrix that stores, for each user and item, the $fRev_{vi}$ feedback received by the reviews provided by $v$;
\item
A $U X U$ matrix that stores the social relations among users.
\end{itemize}

\subsection{Multi-faceted Trust in Collaborative Filtering}
\subsubsection{User-to-User Collaborative Filtering (U2UCF)}
Collaborative recommenders based on U2UCF \cite{Desrosiers-Karypis:11} assume that people who agreed in the evaluation of items in the past are likely to agree in the evaluation of future items.
Let $U$ be the set of users, $I$ the set of items and $R \in {\rm I\!R}^{U X I}$ the users-items rating matrix, where each value is a rating $r_{ui}=R[u,i]$ given by $u \in U$ to $i \in I$. 
U2UCF estimates $u$'s rating of $i$ ($\hat{r}_{ui}$) as follows:
        \begin{equation}
        \label{eq:rmeancenteringcf}
        \hat{r}_{ui} = \bar{r}_u + \frac{ 
        	\sum\limits_{v\in N_i(u)} w_{uv} (r_{vi} - \bar{r}_v)
        }{
        	\sum\limits_{v\in N_i(u)}|w_{uv}|}
    \end{equation}
where $N_i(u)$ is the set of neighbors of $u$ who rated item $i$ and $w_{uv} = \sigma(u,v) \in [0, 1]$ is the observed rating similarity between $u$ and a user $v \in N_i(u)$. 
In U2UCF, $\sigma(u,v)$ is computed by applying a distance metric, e.g., Pearson similarity, between ratings vectors.
\begin{table*}[]
\centering
\caption{Statistics about the Filtered Yelp and LibraryThing Datasets}
\begin{tabular}{lccccc}
\textbf{Yelp}  & \textbf{Min} & \textbf{Max} & \textbf{Mean} & \textbf{Median} & \textbf{Mode} \\ 
\hline
\#Elite years of individual user profiles                                     & 0            & 1            & 0.4296        & 0               & 0             \\
\#Compliments (more+thx+gw) received by individual user profiles     & 0            & 56184        & 83.2143       & 4               & 0             \\
\#Fans of individual users                                             & 0            & 1962         & 13.8254       & 3               & 0             \\
\#Appreciations (useful+funny+cool) on the reviews provided by individual users  & 0            & 43862        & 198.9471      & 59              & 25            \\
\#Appreciations (like) on tips provided by individual users                    & 0            & 966          & 0.4176        & 0               & 0             \\
\#Appreciations (useful+funny+cool) received by individual reviews & 0            & 2278         & 3.9897        & 2               & 0          \\
\#Friends of individual users & 0 & 1887 & 24.2488 & 5 & 1 \\
\hline
\textbf{LibraryThing} & \textbf{Min} & \textbf{Max} & \textbf{Mean} & \textbf{Median} & \textbf{Mode} \\
\hline
\#Appreciations (nhelpful) on the reviews provided by individual users & 0 & 2956 & 19.5842 & 4 & 0  \\
\#Appreciations (nhelpful) received by individual reviews & 0 & 332 & 0.2091 & 0 & 0\\
\#Friends of individual user profiles & 0 & 266 & 3.0958 & 0 & 0
\end{tabular}%
\label{t:feedback}
\end{table*}

\subsubsection{Multi-faceted Trust-based Recommender (MTR)}
This is a variant of U2UCF which identifies neighbors and tunes their impact on rating estimation by applying a general influence metric based on global reputation, local trust and/or observed rating similarity. 
Let $u \in U$, $i\in I$, and $V_i$ the set of users who rated item $i$.
We define the {\em influence of $v$ on $u$ in the context of item $i$} (i.e., how strongly we expect that $v$ conditions $u$ in the evaluation of $i$) as: 
    \begin{equation}
        infl_{uvi}=\beta * \sigma(u,v) + (1-\beta)*t_{uvi}
        \label{eq:beta}
    \end{equation}
where $\sigma(u,v)$ represents the similarity between $u$ and $v$, $t_{uvi}$ is the trust that $u$ has in $v$ in the context of item $i$ (computed by applying Equation \ref{eq:hybridReputation}), and $\beta$ takes value in interval [0, 1]; see below. The set of neighbours of $u$, $N_i(u)$, contains the $n$ users in $V_i$ having the highest values of $infl_{uvi}$.

Given $N_i(u)$, we estimate $u$'s rating of item $i$ by replacing $w_{uv}$ with $infl_{uvi}$ in Equation \ref{eq:rmeancenteringcf}:
       \begin{equation}
        \label{eq:MTR}
        \hat{r}_{ui} = \bar{r}_u + \frac{ 
        	\sum\limits_{v\in N_i(u)} infl_{uvi} (r_{vi} - \bar{r}_v)
        }{
        	\sum\limits_{v\in N_i(u)}|infl_{uvi}|}
    \end{equation}
Equation \ref{eq:beta} above combines similarity and trust in a weighted sum and determines their relative impact using the $\beta$ parameter: the smaller is $\beta$, the more important is trust in the computation of $infl_{uvi}$, and {\em vice versa}.
We combine $\sigma(u,v)$ and $t_{uvi}$ by means of an additive function because we want to give a priority to the people similar to $u$, or who are highly trustworthy.
Notice that, in our experiments, we considered other possible combinations: e.g., the product and the minimum of the two measures, both representing the intuition that neighbors should be trustworthy {\em and} similar to the target user. However, we discarded these combinations because they provided poor recommendation accuracy. This was probably caused by the fact that the pool of candidate neighbors to choose from, in a data sparsity situation, was too small. 

The evidence of similarity between $u$ and $v$ can be modeled as the observed rating similarity of the two users. In that case, we map $\sigma(u,v)$ to the Pearson correlation between rating vectors. However, it can also represent other types of evidence, and in particular those related to social proximity, which can be associated to local trust \cite{Tang-etal:13}. In this second case, we map $\sigma(u,v)$ to $rel_{uv}$ of Equations \ref{eq:friends} or \ref{eq:friendsJS}, depending on the type of context that we want to consider; i.e., without, or with social intersection.

\section{Test methodology}
\label{test}
We evaluate our multi-faceted trust model in Top-N recommendation by applying it to the Yelp \cite{Yelp-dataset} and LibraryThing \cite{LibraryThing-Dataset} datasets. 
On each dataset, we compare MTR with the following baselines: 
User-to-User Collaborative Filtering (U2UCF) \cite{Desrosiers-Karypis:11}, which relies on rating similarity to predict ratings;
TrustMF \cite{Yang-etal:17} and LOCABAL \cite{Tang-etal:13}, which combine rating similarity and trust (mapped to friends relations) using Matrix Factorization.

We evaluate Top-k recommendation performance, with k=10, by taking the ratings observed in the dataset as ground truth. For the evaluation we consider the following metrics: Precision, Recall, F1, RMSE, MAE, MRR, Diversity and User Coverage.

Diversity describes the mean intra-diversity of items in the suggestion lists @k; see \cite{Bradley-Smyth:01}. We assume that items are represented as the sets of categories they are tagged with. The intra-diversity of an individual list @k is thus defined as: 
\begin{equation}
\text{intra-diversity@k}={\frac  {\sum _{{i=1}}^{k}\sum _{{j=i}}^{k} (1 - sim(i, j))} {\frac{k*(k+1)}{2}}}
\end{equation}
where $sim(i, j)$ is the Cosine similarity between the lists of categories associated to items $i$ and $j$ in the dataset. 

On each dataset, we test the algorithms by applying a 10-fold cross-validation, after having randomly distributed ratings on folds: we use 90\% of the ratings as training set and 10\% as test set.

We use the the Surprise \cite{Surprise} implementation of U2UCF, the LibRec \cite{LibRec} implementation of TrustMF, and the RecQ \cite{LOCABAL} implementation of LOCABAL. MTR is developed by extending the Surprise library. 
The implementations of LOCABAL and TrustMF only provide RMSE and MAE metrics to evaluate accuracy; therefore, we limit the comparison to these measures. 

\subsection{Dataset 1: Yelp}
\label{sec:YELP}

The Yelp Dataset \cite{Yelp-dataset}  contains information about a set of businesses, users and item reviews.
Each item is associated with a list of tags that can be interpreted as categories to which the item belongs; e.g., an individual restaurant might be associated to ``Restaurants'', ``Indian'' and ``Nightlife'' categories.
Moreover, each item is associated with a list of item ratings, reviews and tips provided by Yelp users. Every user can provide at most one contribution (including review+rating, and possibly tip) on the same item. Item ratings take values in a [1,5] Likert scale where 1 is the worst value and 5 is the best one. 

User profiles can receive different types of endorsements: e.g.,
every year Yelp rewards its most valuable contributors by attributing them the status of Elite users. 
Moreover, each user profile can receive {\em compliments} by other Yelp users; e.g., ``write more'', ``thanks'' and ``good writer''. Similarly, each review can receive appreciations; i.e., ``useful'', ``funny'' and ``cool''. Notice that the dataset reports the number of compliments and appreciations, but not the identities of the users who provided them.

Yelp enables users to establish generic friend relations in order to filter posts in the social network; moreover, it supports more specific fan relations, which grant a direct access to the contributions provided by the followed users. 
The Yelp dataset publishes the friend relations existing between Yelp users but it only provides the number of fans of each user. Therefore, only the former data can be exploited to infer direct trust-alike relations among users.

For our experiments, we filtered the Yelp Dataset on the users who provided at least 20 ratings, and on the items tagged with the Yelp categories\footnote{The full list of Yelp categories is available at \url{https://www.yelp.com/developers/documentation/v3/category_list}.} that are subclasses of Restaurants and Food: e.g., Cafes, Kebab, Pizza, \dots
The resulting dataset contains 26,600 users, 76,317 businesses, 1,326,409 ratings and 645,020 friend relations. Its users-items rating matrix has sparsity = 0.9994 and its (users-users) friends matrix has sparsity = 0.9991.

The higher portion of Table \ref{t:feedback} provides information about the dataset. It can be noticed that the median number of compliments, fans, appreciations, etc. is very low but it reaches high values in some cases: for each type of feedback, the distribution of individuals (users or reviews) has a long tail. 

\begin{table*}[]
\centering
\caption{MTR Configurations. $Pearson(u, v)$ Denotes the Pearson Correlation between the Rating Vectors of $u$ and $v$}
\begin{tabular}{lcccc}
\textbf{ID} & \textbf{Description} & \textbf{$t_{uvi}$} & \textbf{$\sigma(u, v)$} & \textbf{$rel_{uv}$}\\
\hline \\
MTR-U & B+C+D & $\frac {fb_v+fRev_{vi}+rel_{uv}}{3}$ & $Pearson(u, v)$ & Direct friends \\
MTR-S & A+B+C & $\frac{elite_v + lup_v + opLeader_v + vis_v + fb_v + fRev_{vi}}{6}$ & $Pearson(u, v)$ & -\\
MTR-F & A+D & $\frac{elite_v + lup_v + opLeader_v + vis_v + rel_{uv}}{5}$ & $Pearson(u, v)$ & Direct friends\\
MTR-FS & A &  $\frac{elite_v + lup_v + opLeader_v + vis_v}{4}$ & $Pearson(u, v)$  & - \\
MTR-US & B+C & $\frac{fb_v + fRev_{vi}}{2}$ & $Pearson(u, v)$ & -\\
MTR & A+B+C+D & $\frac{elite_v + lup_v + opLeader_v + vis_v + fb_v + fRev_{vi} + rel_{uv}}{7}$ & $Pearson(u, v)$ & Direct friends\\
MTRTrust1 & A+B+C+D & $\frac{elite_v + lup_v + opLeader_v + vis_v + fb_v + fRev_{vi}}{6}$ & $rel_{uv}$ & Direct friends \\
MTRTrust2 & A+B+C+D & $\frac{elite_v + lup_v + opLeader_v + vis_v + fb_v + fRev_{vi}}{6}$ &  $rel_{uv}$ & Direct friends OR JS(u,v) \\
\end{tabular}%
\label{t:strategies}
\end{table*}

We define the following trust indicators:
\begin{itemize}
    \item[(A)] 
    {\em Global feedback on the user's profile}:
    \begin{enumerate}
    \item $elite_v$: 
    we map each year of Elite status to an element of the $Appreciations_v$ set in Equation \ref{eq:fEndors}:
    \begin{equation}
        elite_v = \frac{\#Elite~ years_v}{max_{a \in U}\#Elite~years_a}
    \end{equation} 
    \item
    $lup_v$ (degree of liking of user profile): we map the requests to write more content ({\em More}), thanks ({\em Thx}) and appreciations of users' writing capabilities ({\em Gw} - good writer) to $Appreciations_v$ in Equation \ref{eq:fEndors}:
    \begin{equation}
       lup_v = \frac{|More_v| + |Thx_v| + |Gw_v|} 
        {max_{a \in U} (|More_a| + |Thx_a| + |Gw_a|)}
    \end{equation} 
     \item 
     $opLeader_v$ (opinion leader degree): the number of fans of a user can be interpreted as a global recognition of her/his profile. We thus map fans to the $Appreciations_v$ in Equation \ref{eq:fEndors}: 
        \begin{equation}
        opLeader_v = \frac{|Fans_v|}
        {max_{a \in U} |Fans_a|}
    \end{equation}
    \item
    {\em Visibility} ($vis_v$): starting from Equation \ref{eq:visibility}, we map the compliments directed to $v$'s user profile to $Appreciations_v$ and the reviews ($Rev_v$) and tips ($Tips_v$) on items provided by $v$ to $Contributions_v$:
        \begin{equation}
            vis_v = \frac{|More_v| + |Thx_v| + |Gw_v|}
                    {maxC*(|Rev_v|+|Tip_v|)}
        \end{equation}
      where $maxC = max_{a \in U} (|More_a| + |Thx_a| + |Gw_a|)$
    \end{enumerate}
    \item[(B)] 
    {\em Global feedback on the user's contributions ($fb_v$):}  
        we map $Contributions_v$ to $Rev_v \cup Tips_v$. Moreover, we map 
        \linebreak 
        $Appreciations_x$ (or $Appreciations_y$) to the feedback obtained by the reviews ("useful" - $Usf$; "funny" - $Fun$; "cool" - $Cool$) and tips ("like" - $Like$) provided by $v$ (or by $a$):
        \begin{equation}
            fb_v = \frac{ 
        	    \sum\limits_{x\in Rev_v \cup Tip_v} |Usf_x|+|Fun_x|+|Cool_x|+|Like_x|
            }{max_{a\in U}\sum\limits_{y \in Rev_a\cup Tip_a} |Usf_y|+|Fun_y|+|Cool_y|+|Like_y|}
        \end{equation} 
    \item[(C)]
    {\em Global feedback on $v$'s review of item $i$ ($fRev_{vi}$):} 
    \begin{equation}
       fRev_{vi} = \frac{ 
       |Usf_{rev_{vi}}|+|Fun_{rev_{vi}}|+|Cool_{rev_{vi}}|
        }{max_{a\in U}(|Usf_{rev_{ai}}|+|Fun_{rev_{ai}}|+|Cool_{rev_{ai}}|)}
        \label{fRev_vi}
    \end{equation} 
    \item[(D)] 
    {\em Social relation between $u$ and $v$ ($rel_{uv}$):} we map social links to friend relations among Yelp users, using either direct friends (Equation \ref{eq:friends}) or direct friends + social intersection (Eq. \ref{eq:friendsJS}). 
    \end{itemize}

\begin{table*}[!htb]
\centering 
\caption{Performance@10 on Yelp Dataset. The Best Values Are in Boldface; the Worst Ones Are Strikethrough}
\begin{tabular}{l|c|cccccccc}
\textbf{Algorithms} & \textbf{$\beta$} & \textbf{Precision} & \textbf{Recall} & \textbf{F1} & \textbf{RMSE} &  \textbf{MAE} & \textbf{MRR} & \textbf{Diversity} & \textbf{User Coverage} \\ \hline

\textbf{U2UCF} & - & \st{0.7634} & 0.7381 & 0.7505 & 1.0518 & 0.7823 & 0.7243 & 0.3027  & \bf{0.8263} \\
\textbf{LOCABAL} & - & - & -  & - & 1.063461 & 0.8361 & - & - & - \\
\textbf{TrustMF} & - & - & -  & - & \st{1.1342} & \st{0.876} & - & - & -\\ \hline

\textbf{MTR-U} & 0.1 & 0.7642 & 0.738 & 0.7509 & 1.0486 & 0.7796 & {\bf 0.725} & \st{0.3023}  & 0.8262 \\

\textbf{MTR-S} & 0.1 & 0.7651 & 0.7385 & 0.7506 & 1.0457 & 0.7785 & 0.7247 & 0.3028 & 0.8249 \\
\textbf{MTR-F} & 0.1 & 0.7651 & 0.7391 & 0.7519 & 1.0452 & 0.778 & 0.7249 & 0.303 & 0.825 \\
\textbf{MTR-FS} & 0.1 & 0.7648 & 0.7387 & 0.7515 & 1.0461 &  0.7788 & 0.7244 & 0.303 & 0.825 \\
\textbf{MTR-US} & 0.1 & 0.7638 & \st{0.7372} & \st{0.7503} & 1.0508 & 0.7817 & 0.7243 & \st{0.3023} & 0.8259 \\
\textbf{MTR} & 0.1 & 0.7653 & 0.7388 & 0.7518 & 1.045 & 0.7778 & {\bf 0.725} & 0.3029 & 0.825 \\

\hline
\textbf{MTRTrust1} & 0.1 & 0.7716 & {\bf 0.7406} & 0.7558 & 1.0765 & 0.8134 & \st{0.6671} & {\bf 0.3329} & \st{0.468} \\
\textbf{MTRTrust2} & 0.1 & {\bf 0.782} & 0.7393 & {\bf 0.76} & {\bf 1.0233} & {\bf 0.7667} & 0.7043 & 0.3157 & 0.6877 \\
\end{tabular}%
\label{tab:performance}
\end{table*}

\subsection{Dataset 2: LibraryThing}
\label{sec:LT}
LibraryThing \cite{LibraryThing} publishes information about books and enables users to create their own virtual libraries and to tag books. In this social network, friends relations are used to watch and take inspiration from the libraries created by other people, as well as to visualize the contributions posted by them. Therefore, these relations are similar to the trust relations of social networks like Epinions, which are established with the intent of monitoring the contributions of the preferred reviewers.
The LibraryThing dataset \cite{LibraryThing-Dataset,Zhao-etal:15} contains item reviews with ratings taken from this social network. In the dataset, items are not classified in any specific category because they are all books; ratings take values in a [1, 5] Likert scale where 1 is the worst value and 5 is the best one. 
For each review, the dataset reports the number of ``helpful'' appreciations it received (``nhelpful'' field of the review). Moreover, it publishes the friend relations between LibraryThing users.  

We filtered this dataset to select the users who provided at least 20 reviews and we removed all the reviews which were not associated with any rating. The resulting dataset contains 12,258 users, 349,365 items, 1,148,270 reviews and 37,949 friend relations. The users-items matrix has sparsity = 0.99973 and the friends matrix has sparsity = 0.99975.

The statistics reported at bottom of Table  \ref{t:feedback} show that, with respect to Yelp, this dataset provides a stricter type of social relation among users and a more limited amount of feedback on their contributions. For instance, the median number of appreciations of an individual review is 0 against the 2 of Yelp. Moreover, as the amount of global feedback collected by user profiles only concerns their reviews, users get very few appreciations from the the social network (4 against 59). Friend relations are fewer, as well: the median number of friends is 0 in LibraryThing.

We consider the following trust indicators:
\begin{enumerate}
    \item[(A)] {\em Global feedback on the user's profile:} missing.
    \item[(B)]
    {\em Global feedback on the user's contributions ($fb_v$):}  
        \begin{equation}
            fb_v = \frac{ 
        	    \sum\limits_{x\in Rev_v} nhelpful_x
            }{max_{a\in U}\sum\limits_{y \in Rev_a} nhelpful_y}
        \end{equation} 
    \item[(C)] 
    {\em Global feedback on $v$'s review about item $i$ ($fRev_{vi}$):} 
    \begin{equation}
       fRev_{vi} = \frac{nhelpful_{rev_{vi}}
        }{max_{a\in U} (nhelpful_{rev_{ai}})}
        \label{fRev_vi}
    \end{equation} 
    \item[(D)]
    {\em Social relation between $u$ and $v$ ($rel_{uv}$):} we map social links to friend relations, using either direct friends (Equation \ref{eq:friends}) or direct friends + social intersection (Equation \ref{eq:friendsJS}). 
\end{enumerate}

\subsection{Multi-faceted Trust Configuration}
\label{sec:configurations}

In order to assess the impact of the A, B, C and D evidence classes on recommendation performance, we test MTR on several configurations of Equation \ref{eq:hybridReputation}, setting the weights of the trust indicators either to 1 or 0 in order to include, or exclude, user profile endorsements (class A), global feedback on the user's contributions (B+C) and social relations (D); see Table \ref{t:strategies}. 

In most of the configurations, we map $\sigma(u, v)$ to the observed rating similarity, which is defined as the Pearson Correlation of $u$ and $v$'s rating vectors. Moreover, we map multi-faceted trust ($t_{uvi}$) to a subset of the trust indicators. Furthermore, we compute social proximity either by considering direct friends (Equation \ref{eq:friends}) or direct friends + social intersection (Equation \ref{eq:friendsJS}).

MTRTrust1 and MTrust2 use all the facets of trust in the computation of $t_{uvi}$ but they ignore the observed rating similarity: the former computes $\sigma(u, v)$ as the local trust between users, taking direct friends relations as an indicator of similarity. The latter computes
$\sigma(u, v)$ by considering direct friends plus the social intersection of users. 

We evaluate the performance of the MTR configurations by varying the value of $\beta$ in $[0, 1]$ in order to find the most convenient setting of Equation \ref{eq:beta} in each configuration. In all the experiments the number of neighbors for rating estimation is 50.

\section{Experimental Results}
\label{results}

\subsection{Yelp}
Table \ref{tab:performance} shows the performance results @10 obtained with $\beta = 0.1$. 
The first portion of the table is devoted to the baselines, i.e., U2UCF, LOCABAL and TrustMF. The second portion groups the results obtained by taking the observed rating similarity into account, and by including or excluding different facets of trust. The last portion of the table reports the results concerning the MTR configurations that focus on multi-faceted trust and ignore rating similarity. 
Figure \ref{fig:curve} compares the RMSE of the baselines with those of MTRTrust2, which outperforms the other configurations in accuracy. We omit the other curves for space reasons.
The table shows that:

\begin{figure}[b]
\centering
\includegraphics[width=0.9\linewidth]{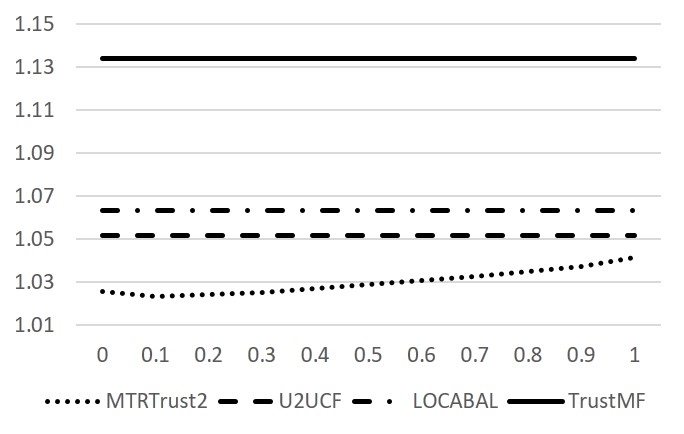}
 \caption{RMSE of the Baselines and of MTRTrust2.}\label{fig:curve}
\hfill
\end{figure}

\begin{itemize}
    \item
    MTRTrust1 and MTRTrust2 are the best algorithms in all the measures except for MRR and User Coverage. 
    \item
    Specifically, MTRTrust2 beats the baselines in all the measures except for User Coverage, which is about 69\% against the 83\% of U2UCF. Indeed, MTRTrust2 obtains the best results with $\beta=0.1$, i.e., by applying $90\%$ global reputation and $10\%$ local trust in Equation \ref{eq:beta}. When $\beta$ increases, performance decreases, but it is always higher than that of the baselines.
    \item 
    MTRTrust2 outperforms MTRTrust1 except for Recall and Diversity and has much higher User Coverage than
    MTRTrust1. This shows that the extension of neighborhood to the social intersection of users, combined with global reputation, helps finding good neighbors for preference estimation. In comparison, the fact that MTRTrust1 focuses on direct friends challenges neighbor identification because it reduces the set of users to choose from.
    \item
    MTRTrust2 also outperforms MTR, which combines global reputation, local trust and observed rating similarity. It thus appears that, when rich trust evidence is available, this is more effective than observed rating similarity in preference prediction.
    Notice also that LOCABAL obtains better results than TrustMF. We believe that this is due to the fact that LOCABAL fuses rating similarity with local trust and global reputation (PageRank score), while TrustMF only uses the first two components.
    \end{itemize}
    We now focus on the second portion of Table \ref{tab:performance} in order to analyze the relative influence of the facets of trust on recommendation performance. Notice that, while $\beta$ increases, these strategies converge to U2UCF because they give progressively more importance to rating similarity. 
    \begin{itemize}
    \item 
    Lines MTR-U, MTR-S and MTR-F compare the performance of the MTR configurations which ignore a single type of trust indicator ($\beta = 0.1$). 
    \begin{itemize}
        \item 
        MTR-U has the worse accuracy and Diversity, showing that, in this dataset, the global feedback on user profiles (class A) strongly influences performance. However, MTR-U has the best user coverage. 
        \item
        Social relations (D) seem to influence performance in a weaker way than user profile endorsements. The minor influence of this type of information can be explained with the fact that, as previously discussed, friend relations provide trust-alike evidence. 
        \item 
        Finally, the feedback on contributions (B+C) appears to be the less influential type of information among the three. 
    \end{itemize}
    \item 
    Consistently with the previous findings, MTR-US has lower results than MTR-U (in some cases, the worst of all algorithms in the table); moreover, MTR-FS performs worse than MTR-F, showing that social relations (D) positively influences recommendation. However, in this dataset, the omission of social relations has minor consequences on accuracy than that of user profile endorsements.  
    \end{itemize}

\subsection{LibraryThing}
Table \ref{tab:performance2} shows the accuracy results @10 with $\beta = 0.1$ on the LibraryThing dataset and it summarizes the values obtained on the Yelp one.
The results are rather different on the two datasets: LOCABAL is the most accurate algorithm when applied to LibraryThing. Moreover, TrustMF has the second best RMSE while U2UCF and MTR-FS (which in this dataset coincides with U2UCF) have the second best MAE. MTRTrust2 obtains the worst accuracy results, which slightly improve when $\beta$ increases (not shown for space reasons), but are always higher than those of the other algorithms. We explain these results as follows:
first, the friend links of this dataset can be interpreted as explicit trust relations concerning user preferences because they are aimed at content sharing. Users establish friend relations in order to view the other users' virtual libraries and contributions. Thus, it would not make sense being friend with somebody who has very different interests than our own. On the other hand, the dataset provides poor feedback on user contributions, and lacks user profile endorsements; therefore, the global reputation deriving from users' feedback is weak and noisy. In this context, LOCABAL is accurate because it combines observed rating similarity with local and global trust inferred from explicit trust relations. In comparison, TrustMF has lower performance because it does not take global reputation derived from trust links into account.
Second, MTR, which fuses observed rating similarity with multi-faceted trust information, has lower accuracy than U2UCF because U2UCF focuses on observed rating similarity that, in this dataset, is more reliable than global reputation derived from anonymous feedback on contributions.
Third, MTRTrust1 and MTRTrust2 have low accuracy because they attempt to compute global reputation on the basis of feedback on users' reviews but the available data is not enough; moreover, they overlook rating similarity, which would help to contrast this lack of information.

\begin{table}[t]
\centering 
\caption{Accuracy@10 on Yelp and LibraryThing Datasets}
\begin{tabular}{l|c|c|c|c|c}
 &  & \multicolumn{2}{c|}{\textbf{Yelp}} & \multicolumn{2}{c}{\textbf{LibraryThing}} 
\\ \hline
\textbf{Algorithms} & \textbf{$\beta$} & \textbf{RMSE} &  \textbf{MAE} & \textbf{RMSE} & \textbf{MAE} \\ \hline

\textbf{U2UCF} & - & 1.0518 & 0.7823 & 0.9214 & 0.6982 \\
\textbf{LOCABAL} & - & 1.063461 & 0.8361 & {\bf 0.8831} & {\bf 0.6817} \\
\textbf{TrustMF} & - & \st{1.1342} & \st{0.876} & 0.9109 & 0.7028 \\
\hline
\textbf{MTR-F} & 0.1 & 1.0452 & 0.778 & 0.9221 & 0.6986
 \\
\textbf{MTR-FS} & 0.1 & 1.0461 & 0.7788 & 0.9214 & 0.6982
 \\
 \textbf{MTR-US} & 0.1 & 1.0508 & 0.7817 & 0.9265 & 0.702
 \\
\textbf{MTR} & 0.1 & 1.045 & 0.7778 & 0.9262 &
0.7017 \\
\hline
\textbf{MTRTrust1} & 0.1 & 1.0765 & 0.8134 & 0.9222 & 0.7107 \\
\textbf{MTRTrust2} & 0.1 & {\bf 1.0233} & {\bf 0.7667} & \st{0.9299} & \st{0.7141} \\
\end{tabular}%
\label{tab:performance2}
\end{table}

\subsection{Discussion}
We previously pointed out that the semantics of friend relations, and the presence of different types and amount of feedback about users, can explain the different performance of the analyzed algorithms. 
In order to further investigate this aspect, we consider a variant of U2UCF (denoted as U2USocial) that replaces rating similarity with social proximity. In U2USocial, the weight $w_{uv}$ of U2UCF in Equation \ref{eq:rmeancenteringcf} is computed as $rel_{uv}$ of Equation \ref{eq:friendsJS}; i.e., it depends on direct friends and social intersection. Basically, U2USocial is similar to MTRTrust2 but it excludes global reputation. 

When applied to Yelp, U2USocial obtains RMSE = 1.0401 and MAE = 0.7748; i.e., it is the second most accurate algorithm, showing that social relations are strong preference predictors. U2USocial is however less accurate than MTRTrust2, which employs both social relations and global reputation derived from user profile endorsements and feedback on users' contributions.
Conversely, on LibraryThing, U2USocial has RMSE = 0.9285 and MAE = 0.713. These are low accuracy values, but they are better than those obtained by MTRTrust2, showing that global reputation derived from feedback on contributions brings noise in preference prediction. As already discussed, the algorithms that combine observed rating similarity with explicit trust links perform better than the pure trust-based ones on this dataset.

The diversity of results obtained by the same algorithms on different domains shows that there is no absolute winner and it highlights the importance of analyzing the characteristics of a dataset before choosing a recommendation algorithm to be applied to it. 

\section{Conclusions}
\label{conclusions}
This paper described a multi-faceted trust model that integrates local trust, represented by social links, with various types of global trust evidence, such as user profile endorsements and feedback on user contributions in social networks.
In order to test the impact of our model on recommendation performance, we integrated it into Collaborative Filtering and we tested the resulting system on two public datasets. We compared the achieved accuracy with that of User-to-User Collaborative Filtering, as well as to that of the LOCABAL and TrustMF trust-based recommenders. The experimental results show that, depending on the characteristics of the dataset, and in particular on the type of trust feedback it provides, the same algorithms can perform rather differently. Before choosing a recommendation algorithm for a specific application domain, it is therefore very important to analyze the characteristics of the domain and of the available types of information about user behavior. 
In our future work, we will evaluate our model on other datasets, possibly considering further facets of trust, in order to validate the results we obtained and to better understand which characteristics influence recommendation performance. Moreover, we will test other weighting methods to combine the facets of trust, rather than only including or excluding them.  

This work was supported by the University of Torino through projects ``Ricerca Locale'', MIMOSA (MultIModal Ontology-driven query system for the heterogeneous data of a SmArtcity, ``Progetto di Ateneo Torino\_call2014\_L2\_157'', 2015-17)   and the Computer Science PhD program.

% REFERENCES FORMAT
% References must be the same font size as other body text.
\bibliographystyle{ACM-Reference-Format}
%\bibliography{plan-rec,httpbib,mybib,sample}
\balance

\end{document}